\documentclass[11pt]{article}
\usepackage{moriond}

\def\be{\begin{equation}}
\def\ee{\end{equation}}
\def\bea{\begin{eqnarray}}
\def\eea{\end{eqnarray}}

\newcommand{\Photo}{}

\begin{document}
\vspace*{4cm}
\title{QCD AND HIGH ENERGY INTERACTIONS: THEORY SUMMARY}

\author{Thomas Gehrmann }

\address{Department of Physics, University of Z\"urich, Winterthurerstrasse 190, 8057 Z\"urich, 
Switzerland\vspace{2mm}}

\maketitle\abstracts{
This article summarizes new theoretical developements, ideas and results that were 
presented at the 2014 Moriond "QCD and High Energy Interactions". }

\section{Introduction: Particle physics after the Higgs boson discovery}
One year ago, at Moriond 2013,
 the ATLAS and CMS collaborations presented 
 for the first time their results on the properties of a 
newly discovered boson~\cite{higgs}. 
A variety of different measurements shown then 
identified the boson to be the long-sought Higgs boson, the mediator of electroweak 
symmetry breaking. This discovery was crowned by the award of the 2013 
Nobel Prize in Physics to Francois Englert and Peter Higgs. 

During the past year, analysis of the LHC7 and LHC8 data continued, providing a wealth of 
new insights into the Higgs boson properties and many other observables. Particle physics is 
also eagerly preparing the next run of the LHC at higher collision energy, which will open up new 
mass ranges in searches for physics beyond the Standard Model. Planning 
for the next generation of high-energy particle colliders has gained a lot of momentum 
throughout the last year. Progress in particle physics relies on a 
fruitful interplay of experiment and theory, 
illustrated in a very lively manner throughout this conference. 

A very prominent example for this interplay is the recent indirect measurement 
of the Higgs boson width, reported by the CMS collaboration~\cite{hwid1}. The 
measurement is based on a comparison of the on-peak and off-peak cross sections for 
four-lepton production, which was suggested only very recently~\cite{hwid2}.

\section{Hadronic physics and QCD at strong coupling}
QCD has been established as theory of strong interactions though a multitude of 
experimental validations. In the high-energy regime, QCD is asymptotically free and 
can be handled with methods of perturbation theory, thereby allowing for 
highly precise theory predictions to be confronted with experimental data. In the low-energy 
regime, QCD becomes strongly coupled, and develops confinement. Quantitative predictions 
in this regime can be obtained only with non-perturbative methods, and are often restricted to 
model systems. 

A commonly used model to study non-perturbative dynamics is N=4 Super-Yang-Mills theory, 
where symmetries and dualities can be used to obtain results at strong 
coupling~\cite{papathanasiou,teramond}. 
In this framework, the pomeron intercept has been computed~\cite{raben,costa,djuric}, 
and model results have been compared to HERA data
from collisions at small $x$. It is observed that the qualitative 
behaviour of the intercept changes substantially between N=4 SYM and QCD. 

An important application of QCD at strong coupling are exclusive decays of heavy quarks, 
which provide an excellent indirect probe of physics beyond the Standard Model at the highest 
energy scales~\cite{petrides}. 
New results on decays to tensor mesons have been derived~\cite{lu}, requiring an extension of the 
effective hamiltonian to include tensor modes. The physics of quarkonium bound states  can be 
described perturbatively in non-relativistic QCD.
 Most recently, this theory has been applied successfully to 
describe the production and polarization of $Y$ mesons~\cite{wan}. 

Heavy-ion collisions probe the non-perturbative dynamics of strong interactions at high density. 
Their theory description must account for a multitude of effects from different areas of classical and 
quantum physics. Many collective effects can be described through models based on 
hydrodynamics~\cite{werner}. To describe parton propagation in heavy-ion collisions, 
the interactions of partons with a surrounding plasma and its excitations have been 
computed~\cite{czajka,mrowczynski}. The high-density regime probed in heavy-ion collisions 
may allow to study phenomena equally relevant to proton interactions at very high energies, such 
as saturation and geometrical scaling~\cite{praszalowicz}. 

The non-perturbative bound state dynamics of quarks and gluons inside the proton enters 
into perturbative predictions of high-momentum transfer observables through the parton 
distribution functions. For most applications, these  functions are 
considered inclusive in transverse momentum, described in the well-established 
framework of collinear factorization and DGLAP evolution equations~\cite{voica,cteq}.  
Several new features appear if the dependence of parton distributions on transverse momentum 
is considered, as required for example for transverse momentum resummation.  Owing to the 
helicity structure of the splitting process, gluons entering an unpolarized 
hadronic collision are linearly 
polarized in the direction of their transverse momentum. Consequently, unpolarized collisions 
at the LHC allow to probe certain spin observables, considered up to now only 
in polarized collisions for example at RHIC. The most promising probe of transverse gluon polarization
at the LHC may be through asymmetries 
in $Y+\gamma$ production~\cite{pisano}, and possible applications of transverse gluon polarization 
could be in probes of the spin and parity of the Higgs boson in decays other than the four-lepton 
channel~\cite{dunnen}. 

\section{Describing collisions at the LHC}
The interpretation of data from particle colliders relies on a close interplay between theory and 
experiment. Precision calculations of collider observables are mandatory for the measurement of 
fundamental parameters, such as masses and coupling constants~\cite{enterria}, for the 
determination of particle properties and for the extraction of auxiliary quantities such as parton 
distributions. Reliable predictions for anticipated signals and their Standard Model backgrounds are
crucial in the design of searches for new physics effects, and in the interpretation of the 
search results in terms of exclusion limits or discovery evidence. 

Theoretical predictions are obtained in perturbation theory, which is truncated at a certain order 
(LO, NLO, NNLO, $\ldots$). The conventional procedure to estimate the error on these
predictions proceeds through the variation of renormalization and factorization scales 
 around a predefined value related to the dominant kinematical 
properties of the process under consideration. The interpretation of this error, its 
combination with other sources of error and its propagation in an analysis of experimental data 
are however questionable. A new approach to the treatment of theoretical errors 
based on Bayesian statistics is currently under development~\cite{jenniches}.

The default standard for theoretical predictions in collider physics are multi-purpose 
simulation programs, usually based on LO calculations augmented by leading-logarithmic 
resummation in the form of a parton shower (PS). Owing to rapid developments in the automation of 
NLO calculations, a new standard is currently emerging in the form of simulation programs 
combining NLO+PS. 

The description of multi-particle production processes often demands a complicated interplay 
of fixed-order descriptions and resummation. In the high-energy limit, the dominant dynamics 
of multi-particle production is described by the BFKL equation. Using this equation, one 
obtains compact approximate forms for high-multiplicity matrix elements, which take proper 
account of multiple large-angle emission~\cite{hapola,matera}. This approach is particularly 
relevant for observables that are poorly described by parton shower calculations, which 
are based on a resummation of small-angle emissions. 

NLO calculations of high-multiplicity observables have seen an enormous progress during the 
past years, especially in view of their automation. This process has been catalyzed by 
the standardization of interfaces between different calculational ingredients (virtual and real 
contributions) in the Binoth Les Houches accord~\cite{blha}. With this, new algorithmic 
developments for the evaluation of one-loop virtual amplitudes can be readily applied to 
physical processes by using established tools and infrastructures for the real radiation as well 
as for event-handling, final state reconstruction and comparison to data. The current frontier in 
multiplicity is set by the calculation of NLO corrections~\cite{kosower1}
 to $W+5j$ production using virtual 
corrections evaluated with the Blackhat package in the Sherpa event generator framework.
 A new development 
in this context is the dissemination of full event information resulting from NLO calculations in 
the form of n-tuples~\cite{kosower2}, 
which can be analyzed subsequently with appropriate final state 
definitions as used in the experimental analysis. 

The combination of NLO calculations with parton showers is by now an established and 
widely used tool. To improve upon this, first steps are now being made to combine NNLO  
calculations with parton shower approximations. A first application in this context is to inclusive 
Higgs production~\cite{re}, 
obtained by combining an NLO+PS description of Higgs+jet production with 
a dedicated scale setting, and a normalization to the inclusive NNLO result. 

Resummation beyond leading logarithms becomes important for 
many precision observables in kinematical situations that receive comparable contributions 
from several different partonic multiplicities. The resulting uncertainties are difficult to quantify. 
The conventional determination of theory errors from scale variations on fixed-order calculations 
only regards a particular multiplicity, and may therefore miss out on the relevance of 
multiple emissions.  A new approach towards 
quantifying theory uncertainties, the efficiency method, 
has been put forward to take proper account of 
kinematical situations prone to large resummation corrections~\cite{zanderighi}, and has been 
applied to Higgs production with a jet veto. 

\section{Precision calculations at NNLO and beyond: results and methods}
Corrections beyond NLO are needed for 
the interpretation of benchmark observables (usually low-multiplicity processes), 
which are measured experimentally to the per cent level as well as for observables with 
potentially large perturbative corrections. For hadron collider processes, a fully differential 
calculation of the higher order corrections, allowing to take into account experimental 
selection criteria and kinematical limitations and thus predicting fiducial cross sections, is 
very much demanded. This can be accomplished by a parton-level event generator, which 
supplies all partonic subprocess contributions with their full kinematical dependence. These 
calculations at NNLO accuracy face two major challenges: the derivation of 
the relevant two-loop matrix elements and the treatment of real radiation corrections. Major progress 
has been made on both aspects in the recent past, and has enabled NNLO calculations for 
several key processes at the LHC. 

Transverse momentum distributions 
of colourless final states at hadron colliders display a universal behaviour 
at low $q_T$, which is well-understood from resummation. This feature is exploited in the 
$q_T$-subtraction technique to extract the singular real radiation contributions, and 
to construct parton-level event generator programs to NNLO accuracy. Recent 
applications of this technique are vector boson pair production~\cite{grazzini} and 
associated $VH$ production~\cite{ferrera}, including the full decay information and QCD 
corrections to the Higgs decay to bottom quarks. 

The LHC experiments measure the top quark pair production cross section to very high precision. 
A similar level of accuracy is now obtained at the theory level with the recent calculation of NNLO 
corrections to the inclusive cross 
section~\cite{fiedler1}, further improved by the inclusion of NNLL resummation 
terms. In the context of the calculation of the two-loop virtual corrections to top quark pair production, 
analytical results for the matching coefficients at 
NNLL~\cite{fiedler2} could be derived as a by-product, again illustrating the fruitful 
interplay of techniques for 
resummation and fixed order calculations at high perturbative orders. 

Calculations multi-loop amplitudes in quantum field theory face major challenges due to the 
large number of amplitudes and Feynman integrals and due to the complicated 
analyticity structure of 
the individual multi-loop integrals. 
By systematically exploiting  relations among integrals, one  obtains a reduction to a minimal set 
of so-called master integrals. To compute these master integrals, various methods have been 
developed, often circumventing the direct integration over the loop momenta~\cite{lee}. A 
particularly powerful technique exploits differential equations in masses and external 
invariants to compute master integrals. This technique has been recently 
systematized~\cite{henn}, 
and currently being  applied to an increasing number of processes. 

Gluon fusion is the dominant Higgs boson production process at the LHC. It receives 
large perturbative corrections at NLO and NNLO, and the determination of the Higgs boson properties 
could in the long run be limited by the theory precision on the NNLO calculation. The calculation 
of N$^3$LO corrections to Higgs production opens up a new era in complexity, with typically a 
hundred-fold increase in the number of diagrams and integrals. A first result in this endeavour is 
the threshold contribution to N$^3$LO, 
which is obtained analytically~\cite{mistlberger}
using the soft-virtual approximation. This calculation makes extensive use of modern 
analytical developments, and establishes many of the technical tools that will pave the way 
towards the derivation of the full N$^3$LO coefficient function.

\section{Going beyond the Standard Model}

Although the Standard Model of particle physics is very successful in describing a wealth of 
experimental data that were taken at collider experiments during the past decades, it is
not considered as ultimate theory of elementary particles, but rather as an effective theory describing 
dynamics at energy scales currently accessible, and possibly above. Searches for 
physics beyond the Standard Model~\cite{harnik} are a primary objective of particle 
physics, both theoretically and experimentally.

Direct searches for the production of new physics signatures 
have produced no evidence for new particles so 
far. They are constrained by the available collider energy, and expectations for next year's LHC run 
at highest-ever collision energy are consequently very high. Indirect 
searches~\cite{petrides} use precision 
observables, rare processes or the internal consistency of the Standard Model. They are not limited 
by the collider energy, but rather by experimental and theoretical precision, available 
luminosity, and last but not least, by the imagination and creativity of particle physicists. 

Using the renormalization group equations, the parameters of the 
Standard Model Lagranigian can be extrapolated to very high 
energies. Of particular interest are the parameters 
related to the Higgs sector, since they determine the form of the effective Higgs potential. Depending 
on the form of the potential, the electroweak vaccum is either stable (absolute minimum of the 
potential), metastable (existence of a lower minimum at larger field values) or unstable 
(potential unbounded from below at large field values). The effective Higgs potential is particularly 
sensitive on the measured masses of the Higgs boson and the top quark, and current data point 
towards a metastable situation. In the case of metastability, tunnelling from the electroweak 
vacuum to the vacuum at large field values is possible. The tunnelling rate computed in the 
Standard Model (i.e.\ in absence of any physics beyond the Standard Model, even at the Planck 
scale) is however many orders of magnitude larger than the age of the universe.  The tunnelling process 
is mediated by extended field configurations (instantons). Even in absence of 
physics beyond the Standard Model, it provides sensitivity to dynamics around the  Planck scale, 
since the typical instanton size is only about a factor ten larger than the Planck 
length. Following this observation, it has recently been 
demonstrated~\cite{branchina} that new physics at the Planck scale could lead to much smaller 
tunnelling times, thus invalidating the metastability condition. These new insights may provide 
very valuable input to model building for Planck scale physics. 

One of the strongest motivations for physics beyond the Standard Model is the 
astrophysical evidence for dark matter in galaxies and in cosmological observations. Besides 
its gravitational effects, one expects dark matter to interact only weakly with known matter, and 
scenarios for dark matter production in the early universe favour masses of dark matter particles 
of the order of the weak scale. 
Direct searches for dark matter in recoil experiments have not produced conclusive evidence so 
far, and many realizations for weakly interacting massive particles are now constrained rather severely. 
An alternative scenario is provided by the so-called Higgs portal models, where dark matter and 
ordinary matter couple only through Higgs interactions, resulting in considerably lower rates for 
direct detection. A possible realization of the Higgs portal 
scenario is the inert doublet model~\cite{swiezewska,yu}, which will be probed at the next generation 
of direct detection experiments and which predicts small but detectable deviations in the Higgs boson 
properties. 

Indirect searches for new physics in certain flavour observables are now reaching well beyond the 
energy scales probed in direct searches at the LHC. Visible, but not yet significant, 
discrepancies between experimental 
measurements and theoretical expectations are observed especially in $B\to K l^+l^-$ final 
states~\cite{petrides} and in $B\to \tau \nu$~\cite{crivellin}. Both types of anomalies motivate 
theoretical speculation on possible new physics in the flavour sector, and further 
experimental validation is expected in due course.

\section*{Acknowledgments}
I would like to thank the organizers of Moriond 2014 for creating a very unique 
environment for discussion and interchange of ideas, and all participants of this meeting for 
a multitude of new insights. My work is supported in part by  
the Swiss National Science Foundation (SNF) under contract 200020-149517 
and by the Research Executive Agency (REA) of the European Union under the Grant Agreements
 PITN-GA-2010-264564 (LHCPhenoNet), PITN-GA-2012-316704 (HiggsTools), 
 and the ERC Advanced Grant MC@NNLO (340983).

\end{document}